\documentclass[doublecol]{epl2} 

\usepackage{amsmath}
\usepackage{amssymb}

\title{Conformal invariance: from Weyl  to $SO(2,d)$ }
\shorttitle{} 

\author{Sofiane Faci \thanks{E-mail: \email{sofiane@cbpf.br}}}

\institute{                    
Institute of Cosmology, Relativity and Astrophysics (ICRA - CBPF),
Rua Dr. Xavier Sigaud, 150, Urca, CEP 22290-180, Rio de Janeiro, RJ, Brasil
}
\pacs{11.25.Hf}{Conformal field theory, algebraic structures.}
\pacs{04.62.+v}{Quantum fields in curved spacetime}
\pacs{02.20.Qs}{General properties, structure, and representation of Lie groups}

\abstract{
The present work deals with  two different but subtilely related kinds of conformal mappings: Weyl rescaling in $d>2$ dimensional spaces and $SO(2,d)$ transformations. We express how the difference between the two can be compensated by diffeomorphic transformations. 
This is well known in the framework of String Theory but in the particular case of $d=2$ spaces. Indeed, the Polyakov formalism describes world-sheets in terms of two-dimensional conformal field theory.
On the other hand, B. Zumino had shown that a classical four-dimensional Weyl-invariant field theory restricted to live in Minkowski space leads to an $SO(2,4)$-invariant field theory. 
We extend Zumino's result to relate Weyl and $SO(2,d)$ symmetries in arbitrary conformally flat spaces (CFS). 
This allows us to assert  that a classical $SO(2,d)$-invariant field does not distinguish, at least locally, between two different $d$-dimensional CFSs. 
}

\begin{document}
\maketitle

Weyl rescalings and $SO(2,d)$ transformations have different natures, the first is a metric rescaling and the second is a coordinate system transformation.  However, the two are subtilely related. They are both conformal transformations and do preserve the space causal structure. As a consequence, some confusion can appear when dealing with the two transformations at the same time. Our purpose is to clarify this confusion by giving the explicit link between them.
The conformal symmetry is a fundamental ingredient for theoretical physics covering several areas \cite{Kastrup}.  
A huge amount of papers deal with Weyl invariance, among which we cite \cite{ORaifeartaigh:2000, Scholz:2011fk, Scholz:2011uq, Castro:2009zza, Gover:2008-tractors, Maki:2012fk, Romero:2012uq}. The same thing applies for CFT in  $d>2$ spaces \cite{Maldacena:1998, Maldacena:2011jn, Antoniadis:2011, Antoniadis:2011rr, Nikolov:2007, Todorov:2012xx, Costa:2010, Costa:2011} and $d=2$ \cite{cft-2d, Polyakov:1970, Itzykson:1988}.

B. Zumino had shown how Weyl invariance  implies $SO(2,4)$ invariance in a flat space   \cite{Zumino}:
\textit{
If a classical four-dimensional field theory is invariant under Weyl and diffeomorphic transformations, its restriction to Minkowski space is $SO(2,4)$-invariant.} 

In other words, the difference between the two conformal transformations can be,  in this case, compensated by diffeomorphisms, which was earlier indicated in \cite{Fulton}.
This result is important since, on one hand, it provides the explicit link between the two kinds of conformal transformations. On the other hand, it allows us to reduce the calculations of the fifteen parameter $SO(2,4)$ group action to calculations under Weyl transformations, much easier. This was exploited, for instance, in the works \cite{Kraus:1992ru, Kraus:1992th, Weinberg:2010, pconf5, pconf6}.
Let us see how it works with a simple example. 
It is easy to check the Weyl invariance of the equation $(\Box + \frac{1}{6}$$  R)\phi=0$ for a scalar field $\phi$ defined in a four-dimensional space. Also, Zumino's statement ensures the $SO(2,4)$ invariance of its minkowskian limit $\partial^2 \phi = \eta^{\mu\nu} \partial_{\mu}\partial_{\nu}\phi=0$. In both cases, Weyl and $SO(2,4)$, the field $\phi$ transforms with the same conformal weight $-1$.

This is well known for the world-sheet  description in the context of the Polyakov formalism of String theory \cite{Polyakov:1987, Friedan:1985ge, Belavin:1984vu, CFT-Strings}. Nevertheless, the latter deals only with two-dimensional CFT, a singular case where the conformal group becomes infinite.
A. Iorio et al.  explored a sorte of the inverse relation  \cite{Iorio:1996}. The authors developed the Weyl-gauging method which extends a minkowskian $SO(2,4)$-invariant field theory to construct a Weyl-invariant theory.
 Yet, the authors did not mentioned Zumino's work and said nothing about $SO(2,d)$ invariance in curved spaces. Afterwards, R. Jackiw showed that the same procedure holds  in two-dimensional spaces  \cite{Jackiw:2005su}.


The present work extends Zumino's statement in the following terms:
\textit{ 
If a classical $d>2$ dimensional field theory is invariant under Weyl and diffeomorphic transformations, its restriction to a conformally flat space (CFS) is  $SO(2,d)$-invariant.} 

\newpage	

The demonstration can be splited in two parts:
\begin{itemize}
\item The restriction of a Weyl- and diffeomorphic-invariant field theory to a d-dimensional Minkowski space leads to an $SO(2,d)$-invariant theory.
\item The transition from the minkowskian $SO(2,d)$-invariant field theory to its corresponding d-dimensional CFS theory is performed by a structure transport using a Weyl rescaling. 
\end{itemize}

The first part of the proof is justified in appendix B, 
which restores for arbitrary $d>2$ the demonstration given by Zumino \cite{Zumino}.
In order to develop the second part, let us work in a $d>2$ dimensional real manifold $\cal M$ and consider an arbitrary classical field $F$ defined first in Minkowski space $({\cal M}, \eta_{\mu\nu})$ with the signature ($+,--\dots$) and obeying to the generic dynamical equation \begin{equation}\label{equation-F}
\mathcal{E}F = 0,
\end{equation}
where $\mathcal{E}$ is a first- or higher-order differential operator.
This equation is assumed to be invariant under the infinitesimal action of $SO(2,d)$.
This is because we are interesting in the local structures, i.e. algebraic structures. Also we considere the associated algebra $so(2,d)$. The invariance of (\ref{equation-F}) under any element $e\in SO(2,d)$ is then expressed as
\begin{equation}\label{invariance-equation}
\forall X_{e} \in so(2,d), \qquad [ {\cal E}, X_{e} ] = \zeta_{e} \, {\cal E},
\end{equation}
where $\zeta_{e}$ are real functions depending on $e$. The $1+\frac{d}{2}(d+3)$ pairs $\{e, \zeta_{e}\}$ formalize the $SO(2,d)$-invariant structure of the theory, see appendix B 
for more details. The set of solutions of the equation (\ref{equation-F}) forms a representation of the $so(2,d)$ algebra. This representation can be expressed by commutation relations 
\begin{equation}\label{action-F}
[X_{e}, F] = X_{e} F + \Sigma_{e} F,
\end{equation}
where $X_{e}F$ denotes the scalar action and $\Sigma_{e}F$ the tensorial (or spinorial) action.



Now, the whole structure  - (\ref{equation-F}, \ref{invariance-equation}, \ref{action-F}) - of the minkowskian $SO(2,d)$-invariant field theory can be transported to a CFS $(\mathcal{M}, \bar g_{\mu\nu})$. The latter is locally related to Minkowski space by a Weyl rescaling 
\begin{equation}\label{weyl}
\bar{g}_{\mu\nu} = K^2 \, \eta_{\mu\nu},
\end{equation}
where the Weyl factor $K$ is a real, non-vanishing and smooth ($C^\infty$) function.
To do so, we define two maps: the map $W$ to transform the fields and the map $H$ to transform the differential operators acting on them, 
\begin{eqnarray}
W : F  & \rightarrow & \bar{F} = W F \label{W}
\\
H : \mathcal{O}  & \rightarrow & \bar{\mathcal{O}} = W \mathcal{O} W^{-1} \label{H},
\end{eqnarray}
where $W$ denotes the map and its matrix representation, it depends on the Weyl factor $K$ and its derivatives.
Usually, $W$ takes the form $WF=K^s\, F$ where $s$ is called conformal weight (or dimension) of the field; but more general definitions can be considered for non scalar fields \cite{pconf3, pconf5}.

First, we use these two maps to transport the equation (\ref{equation-F}) to get the dynamical equation in the CFS
\begin{equation}\label{equation-F-bar}
\begin{array}{ccl}
\bar{\mathcal{E}} \bar F & =& H {\cal E}\ W F
\\
 & =  &   W \mathcal{E} W^{-1} \, W F 
\\
  & =  &   W (\mathcal{E} F)
\\
 & = & 0.
\end{array}
\end{equation} 
Then we trasport the generators $X_{e}\in so(2,d)$ (they are differential operators) by the map $H$
\begin{equation}\label{symm}
X_{e} \to \bar{X}_{e}= H X_{e} = W X_{e} W^{-1}.
\end{equation}
The generators $\bar X_{e}$ form a new basis for the algebra $so(2,d)$
and the invariance of the equation (\ref{equation-F-bar}) is implemented as follows
\begin{equation}\label{structure}
\begin{array}{ccl}
[\bar{\mathcal{E}}, \bar{X}_{e}]   
& =  & [H \mathcal{E}, H X_{e}]
\\
  & = & W [\mathcal{E}, X_{e}] W^{-1}
  \\
  & = & \zeta_{e} \, \bar{\mathcal{E}}.
\end{array}
\end{equation}
That is to say
$
\bar{\zeta}_{e}=\zeta_{e},$
which means that the resulting theory defined in the CFS $({\cal M}, \bar g_{\mu\nu})$ keeps exactly the same $SO(2,d)$-invariant structure of the minkowskian theory.
Finally, the group representation of the symmetry group (\ref{action-F}) is transported as follows
\begin{equation}\label{action-transport}
\begin{split}
[\bar X_{e}, \bar F] & = \bar X_{e} \bar F + \bar \Sigma_{e} \bar F 
\\ & 
= H X_{e} \  \bar F  + H \Sigma_{e}  \ \bar F
\\ & 
= W X_{e} W^{-1} \  \bar F  +  \Sigma_{e}  \ \bar F
\\ & 
= W X_{e} \ F  +  \Sigma_{e}  W \ F
\\ & 
=  W\, [ X, F ].
\end{split}
\end{equation}

Doing so, we have succeeded in formally transport the whole classical structure of a minkowskian $SO(2,d)$-invariant theory to obtain a new $SO(2,d)$-invariant theory defined in an arbitrary d-dimensional CFS.  This demonstrates the statement we have proposed above. This allows us to assert that a classical $SO(2,d)$-invariant field does not distinguish, at least locally, between two different CFSs. 

This result might appear intuitive but, to our knowledge, this was never clearly stated in the literature.
In  \cite{Deser:1983}, Deser and Nepomechie wrote ``Weyl invariance is the essential factor needed for null cone field propagation in a constant curvature space'', and since 
$SO(2,4)$ invariance was also known to be responsible for null cone propagation in constant curvature spaces,
 the authors concluded that Weyl invariance implies $SO(2,4)$ invariance in such spaces. They pointed out in a footnote ``Presumably, the result can be extended to general conformally flat spaces''.
In addition, a recently published article \cite{Clark:2012ef} explored the link between  Weyl rescalings in four-dimensional spaces and local dilations, one of the fifteen $SO(2,4)$ transformations, without even referring to Zumino's work.

Note that at the stages (\ref{invariance-equation}) and (\ref{action-F}) one would be attempted to replace $SO(2,4)$ by any other Lie group $G$ and follow the same steps (\ref{symm} - \ref{action-transport}). But here is the reason why this is not true. A well defined theory in a given space must be invariant under the associated isometry group. This is where the $SO(2,4)$ plays a central role. Indeed, the smallest group containing, as subgroups, all isometry groups associated to the CFSs is exactly the restricted conformal group \cite{Todorov:1978rf, Keane:1999}. 

%
The set of conformally flat spaces is of high importance since it gathers all relevant spaces for cosmology: Minkowski, de Sitter, Anti-de Sitter and FLRW type spaces.
We have used the Weyl equivalence between CFSs and Minkowski space, but a Weyl rescaling is a local transformation and depends on a coordinate system. The latter does not, in general, cover the whole spaces. Nonetheless, several coordinate systems can be used to compensate this defect, examples are given in \cite{Higuchi:2008, pconf4}. Moreover, $SO(2,d)$ is a Lie group which ensures that the global action can be obtained from the infinitesimal one by using the exponential application. Note that some CFSs,  like tori, break the $SO(2,4)$ invariance globally. This difficulty is avoided since we are looking exclusively to the algebraic (local) structure of $SO(2,4)$-invariant field theories. Also, no global behaviors were discussed. 

In this work, only classical fields were considered. Quantum fields are much more subtle to deal with. In particular, the quantization of $SO(2,d)$-invariant fields leads to the so-called conformal anomaly since the renormalization procedure introduces a typical scale (cutoff) which breaks the conformal invariance. The situation is worst in curved spaces, even though these are conformally flat. These questions are important but they are left for future investigations.

\section{Appendix A: $\ SO(2,d)$ transformations}\label{groupe-conforme}
In a $d>2$ dimensional flat space $\eta_{\mu\nu}=(+,--\dots)$, the $SO(2,d)$ group acts as follows
\begin{equation}\label{transformation-conforme}
\begin{split}
& e\in SO(2,d) :\ x^\mu \to e.x^\mu = \bar  x^{\mu}
\\
& \eta_{\mu\nu} d\bar x^\mu d\bar x^\nu  = \omega_{e}^2(x)\, \eta_{\mu\nu} d x^\mu d x^\nu,
 \end{split}
 \end{equation}
 where $\omega_{e}$ is the conformal factor of the element $e$.
Thus, Minkowski space is left invariant; and this holds for any CFS. This is why $SO(2,d)$ is sometimes calles the restricted conformal group. $SO(2,d)$ is a $1+(d+3)d/2$ ($15$ when $d=4$) parameter group corresponding to:
\begin{itemize}
\item
Translations: $d$ parameters $a^\mu$
\item Lorentz transformations: $(d-1)d/2$ parameters $\Lambda^\mu_{\nu}$
\item Dilatation: $1$ parameter $\lambda$
\item SCT: $d$ parameters given by a vector $b^\mu$.
\end{itemize}
The infinitesimal transformations are obtained when the parameters $(a^\mu, \Lambda^{\mu\nu} = \eta^{\mu\nu} + \theta^{\mu\nu}, \lambda = 1+ \epsilon, b^\mu)$ are taken in their infinitesimal limits. The new parameters $\lambda^{A}_{e}=(a^\mu, \theta^{\mu\nu}, \epsilon, b^\mu)$, where $A$ is the $\lambda_{e}$ index, are all infinitesimal. Let us write the variations
\begin{equation}\label{transfo-infinitesimal} 
\delta_{e} x^\mu =  e. x^\mu - x^\mu = \lambda_{e}^{ A} \ \delta_{ A}^{e} x^{ \mu},
\end{equation}
with $  \delta_{\mu}^{ T} x^{\nu} = \delta_{\mu}^{\nu}, $ 
$ \delta_{ \mu\nu}^{ L} x^{\tau} =  \delta_{\mu}^\tau x_{\nu} - \delta_{\nu}^\tau x_{\mu},  
$
$  \delta^{ D} x^{\mu} =  x^\mu and
$
$ \delta^{ S}_{\mu} x^{\nu} =  2 x_{\mu}x^\nu - \delta_{\mu}^\nu x^2.
$
Now, let us define a conformal field $F$ of weight $s$. To get conformal invariance, one needs, for instance, $s=(2-d)/2$ for a scalar field or $s=(4-d)/4$ for a vector field.
The variations of $F$ under (\ref{transfo-infinitesimal}) read $\delta_{e}\, F(x)  = \lambda^{ A}_{e} \  \delta_{ A}^{e} F (x)$ with 
\begin{equation}\label{representation-s}
 \delta_{ A}^{e} F (x) =  \left(\delta_{ A}^{e}x.\partial + \frac{s}{d}\, \partial. \delta_{ A}^{e}x + \Sigma_{ A}^{e} \right) \, F (x),
\end{equation}
where $\Sigma_{ A}^{e}$ denotes the spinorial transformation of the field.
These transformations can be written as commutation relations,
$\delta_{e}\, F(x)  = \lambda^{ A}_{e} \ [X_{ A}^{e}, F]$, where $X^{e} \in so(2,d)$ are the group generators.
One obtains
\begin{eqnarray}
  \delta^{ T}_{\mu} F & = & \partial_{\mu} \, F, \label{translations-F}
 \\
\delta^{ L}_{\mu\nu} F & = & \left( x_{\mu}\partial_{\nu} - x_{\nu}\partial_{\mu} + \Sigma_{\mu\nu}^{ L}  \right) F, \label{Lorentz-F}
\\
 \delta^{ D} F & =  & \left( x.\partial + s \right) F, \label{dilatation-F}
\\
\delta^{ S}_{\mu} F &=&  \Big(  2x_{\mu} x.\partial - x^2\partial_{\mu} + 2s\, x_{\mu} +\Sigma_{\mu}^{ S} \Big) F.\ \ \label{SCT-F}
 \end{eqnarray}
This provides an $so(2,d)$ representation which is $d$-independent but $s$-dependent. Nevertheless, the algebra is $d$- and $s$-independent, it is expressed using the generators $X_{e}$, denoted as usual by $P,M,D$ and $K$ and extracted from (\ref{translations-F}-\ref{SCT-F}):
$  P_{\mu}  =  \ \partial_{\mu} $, 
$M_{\mu\nu}  = \ x_{\mu}\partial_{\nu} - x_{\nu}\partial_{\mu}$,
$ D  =  \ x.\partial + s$ and
$K_{\mu}   = \  2 x_{\mu} x.\partial - x^2 \partial_{\mu}  + 2sx_{\mu}$. The $so(2,d)$ algebra reads
\begin{equation}\label{algebre-conforme}
\begin{array}{rll}
 [M_{\mu\nu} , M_{\lambda\tau}] &=&  \eta_{\mu\tau} M_{\nu\lambda}  - \eta_{\mu\lambda}M_{\nu\tau} 
 +   \eta_{\nu\lambda}M_{\mu\tau} - \eta_{\nu\tau}M_{\mu\lambda} \\
\ [ P_{\mu} , P_{\nu} ] &=& [K_{\mu}, K_{\nu}] = [D, M_{\mu\nu}] = 0\\ 
\ [P_{\lambda} , M_{\mu\nu}] &=& \eta_{\lambda \mu} P_{\nu} -  \eta_{\lambda \nu} P_{\mu} \\
\ [K_{\lambda} , M_{\mu\nu}] &=& \eta_{\lambda \mu} K_{\nu} -  \eta_{\lambda \nu} K_{\mu} \\
\ [P_{\mu}, K_{\nu}] &=& 2 (\eta_{\mu\nu} D - M_{\mu\nu})\\
\ [P_{\mu}, D] &= & P_{\mu}, \quad \ [D, K_{\mu}] = K_{\mu}.
\end{array}
\end{equation}

\section{Appendix B: \ From Weyl to $SO(2,d)$ in Minkowski space}\label{proof}
The demonstration of Zumino's statement for $d>2$ is basically the same as that given by Zumino for $d=4$ \cite{Zumino}.
Let us consider a field $F$ of conformal weight $s$, defined in a d-dimensional spacetime 
$({\cal M}, g_{\mu\nu})$ and verifying the equation
\begin{equation}\label{eq-zumino}
{\cal E}F=0,
\end{equation} 
which is invariant under the (infinitesimal, $K=1+k, k\ll 1$) Weyl transformations (compare with (\ref{weyl}))
\begin{equation}\label{weyl-infinitesimal-bis}
\left \{
\begin{array}{lll}
\delta g_{\mu\nu} & = & 2\ k(x) \ g_{\mu\nu},
\\
\delta F &  = & s \ k(x) \ F.
 \end{array}
\right .
\end{equation}
Furthermore, a well defined physical equation in General Relativity has to be invariant under the diffeomorphic transformations (coordinates transformations)
\begin{equation}\label{diffeo}
\delta x^\mu =  \bar x^\mu - x^\mu = \xi^\mu(x),
\end{equation}
where $ \xi^\mu(x)$ are arbitrary real and infinitesimal functions. This implies the following (Einstein) transformations of the metric tensor and the field $F$
\begin{equation}\label{einstein-infinitesimal}
\left \{
\begin{array}{lll}
\delta g_{\mu\nu} &= & \xi .\partial g_{\mu\nu} + \partial_{\mu} \xi^\lambda g_{\lambda\nu} + \partial_{\nu} \xi^\lambda g_{\lambda\mu} ,
\\
\delta F & = & \xi .\partial F + \Sigma F,
 \end{array}
\right .
\end{equation}
where $\Sigma$ denotes the spinorial action.

We want to show that the restriction of the equation (\ref{eq-zumino}) to Minkowski space ($g_{\mu\nu}=\eta_{\mu\nu}$) is $SO(2,d)$-invariant. To do this, we choose the transformations 
 (\ref{diffeo}) to correspond to those of the $SO(2,d)$ group (\ref{transformation-conforme}).

\textit{Dilations:}\\
Let $\xi^\mu= \epsilon x^\mu$, the variations (\ref{einstein-infinitesimal}) become
 \begin{equation}\label{demo1}
\left \{
\begin{array}{lll}
\delta g_{\mu\nu} &= & \epsilon \, ( x.\partial +2 )g_{\mu\nu},
\\
\delta F & = & \epsilon \ x .\partial F.
 \end{array}
\right .
\end{equation}
For the Weyl transformation (\ref{weyl-infinitesimal-bis}) we choose $k(x)=-\epsilon$. We obtain
\begin{equation}\label{demo2}
\left \{
\begin{array}{lll}
\delta g_{\mu\nu} & = & - 2\ \epsilon \ g_{\mu\nu},
\\
\delta F &  = & - s \ \epsilon \ F.
 \end{array}
\right .
\end{equation}
Summing the two contributions (\ref{demo1}) and (\ref{demo2}), we find
\begin{equation}
\left \{
\begin{array}{lll}
\delta g_{\mu\nu} & = &  \epsilon \ x.\partial \ g_{\mu\nu},
\\
\delta F &  = &  \epsilon\, (x .\partial - s ) \ F.
 \end{array}
\right .
\end{equation}
Finally, imposing $g_{\mu\nu} = \eta_{\mu\nu}$, implies the right dilation transformation (\ref{dilatation-F}).

\textit{SCT:}\\
Following the same steps than for the dilations but this time choosing $\xi^\mu = 2 \, x^\mu b.x - b^\mu x^2 $ in (\ref{einstein-infinitesimal})
 \begin{equation}\label{demo3}
\left \{
\begin{array}{lll}
\delta g_{\mu\nu} &= &  b^\mu \left( 2\, x_{\mu} x.\partial - x^2\partial _{\mu} - 4 x_{\mu} \right) \ g_{\mu\nu},
\\
\delta F & = & b^\mu \left( 2\, x_{\mu} x.\partial - x^2\partial _{\mu} + \Sigma_{\mu}^{ S} \right) \, F,
 \end{array}
\right .
\end{equation}
and taking $k(x)=2\, b.x$ in (\ref{weyl-infinitesimal-bis}) then summing the result with (\ref{demo3}) produces
\begin{equation}
\left \{
\begin{array}{lll}
\delta g_{\mu\nu} & = & b^\mu \left( 2 x_{\mu} x.\partial - x^2\partial _{\mu} \right) \ g_{\mu\nu},
\\
\delta F &  = &  b^\mu \left( 2 x_{\mu} x.\partial - x^2\partial _{\mu} + 2s x_{\mu} + \Sigma_{\mu}^{ S} \right) \ F,
 \end{array}
\right .
\end{equation}
\\
which, imposing $g_{\mu\nu} = \eta_{\mu\nu}$ at the end, implies the right SCT transformation law (\ref{SCT-F}).

\textit{Translations and Lorentz transformations:}\\
It is sufficient to set the functions $\xi^\mu$ of (\ref{diffeo}) to correspond to the translations and Lorentz transformations. These are isometries of Minkowski space and there is no need to involve the Weyl rescaling, so we choose $k=0$. The combination of (\ref{weyl-infinitesimal-bis}) and (\ref{einstein-infinitesimal}) thus corresponds to the right translations and Lorentz transformations of the $SO(2,d)$ group (\ref{translations-F}) and (\ref{Lorentz-F}).


\acknowledgments
The author would like to thank M. Novello for constructive discussions, R. Jackiw for useful references and N. Pinto-Neto for reading the manuscript. This work was supported by CNPq.

\bibliographystyle{eplbib}
\bibliography{Biblio}

\end{document}